\documentclass[twocolumn,preprintnumbers,amsmath,amssymb]{revtex4}

\usepackage{graphicx}
\usepackage{dcolumn}
\usepackage{bm}
\usepackage{units}

\begin{document}

\title{The rapidity and centrality dependence of nuclear modification
  factors at RHIC --- what does bulk particle production tell us
  about the nuclear medium?}

\newcommand{\bnl}           {$\rm^{1}$}
\newcommand{\ires}          {$\rm^{2}$}
\newcommand{\kraknuc}       {$\rm^{3}$}
\newcommand{\krakow}        {$\rm^{4}$}
\newcommand{\baltimore}     {$\rm^{5}$}
\newcommand{\newyork}       {$\rm^{6}$}
\newcommand{\nbi}           {$\rm^{7}$}
\newcommand{\texas}         {$\rm^{8}$}
\newcommand{\bergen}        {$\rm^{9}$}
\newcommand{\bucharest}     {$\rm^{10}$}
\newcommand{\kansas}        {$\rm^{11}$}
\newcommand{\oslo}          {$\rm^{12}$}

\author{
  {\bf \underline{B.~H.~Samset}} 
  for the BRAHMS Collaboration\\
  I. Arsene\bucharest, 
  I.~G.~Bearden\nbi, 
  D.~Beavis\bnl, 
  C.~Besliu\bucharest, 
  B.~Budick\newyork, 
  H.~B{\o}ggild\nbi, 
  C.~Chasman\bnl, 
  C.~H.~Christensen\nbi, 
  P.~Christiansen\nbi, 
  J.~Cibor\kraknuc, 
  R.~Debbe\bnl, 
  E. Enger\oslo,  
  J.~J.~Gaardh{\o}je\nbi, 
  M.~Germinario\nbi, 
  K.~Hagel\texas, 
  H.~Ito\bnl, 
  A.~Jipa\bucharest, 
  F.~Jundt\ires, 
  J.~I.~J{\o}rdre\bergen, 
  C.~E.~J{\o}rgensen\nbi, 
  R.~Karabowicz\krakow, 
  E.~J.~Kim\bnl, 
  T.~Kozik\krakow, 
  T.~M.~Larsen\oslo, 
  J.~H.~Lee\bnl, 
  Y.~K.~Lee\baltimore, 
  S.~Lindal\oslo, 
  G.~L{\o}vh{\o}iden\oslo, 
  Z.~Majka\krakow, 
  A.~Makeev\texas, 
  B.~McBreen\bnl, 
  M.~Mikelsen\oslo, 
  M.~Murray\texas, 
  J.~Natowitz\texas, 
  B.~Neumann\kansas, 
  B.~S.~Nielsen\nbi, 
  J.~Norris\kansas, 
  D.~Ouerdane\nbi, 
  R.~P\l aneta\krakow, 
  F.~Rami\ires, 
  C.~Ristea\bucharest, 
  O.~Ristea\bucharest, 
  D.~R{\"o}hrich\bergen, 
  B.~H.~Samset\oslo, 
  D.~Sandberg\nbi, 
  S.~J.~Sanders\kansas, 
  R.~A.~Scheetz\bnl, 
  P.~Staszel\nbi, 
  T.~S.~Tveter\oslo, 
  F.~Videb{\ae}k\bnl, 
  R.~Wada\texas and 
  Z.~Yin\bergen, 
  I.~S.~Zgura\bucharest\\ 
  \bnl~Brookhaven National Laboratory, Upton, New York 11973 \\
  \ires~Institut de Recherches Subatomiques and Universit{\'e} Louis
  Pasteur, Strasbourg, France\\
  \kraknuc~Institute of Nuclear Physics, Krakow, Poland\\
  \krakow~Smoluchkowski Inst. of Physics, Jagiellonian University, Krakow, Poland\\
  \baltimore~Johns Hopkins University, Baltimore 21218 \\
  \newyork~New York University, New York 10003 \\
  \nbi~Niels Bohr Institute, Blegdamsvej 17, University of Copenhagen, Copenhagen 2100, Denmark\\
  \texas~Texas A$\&$M University, College Station, Texas, 17843 \\
  \bergen~University of Bergen, Department of Physics, Bergen, Norway\\
  \bucharest~University of Bucharest, Romania\\
  \kansas~University of Kansas, Lawrence, Kansas 66049 \\
  \oslo~University of Oslo, Department of Physics, Oslo, Norway\\
 }

\date{\today}

\begin{abstract}

The BRAHMS experiment at RHIC has measured the production of charged
hadrons as a function of pseudorapidity and transverse momentum in
$Au+Au$, $d+Au$ and $p+p$ collisions at a common energy of
$\sqrt{s_{NN}}=\unit[200]{GeV}$, and from these spectra we construct
the nuclear modification factors for both ``hot'' and ``cold'' nuclear
matter.

In this contribution I will show how these factors evolve with
pseudorapidity and collision centrality. We see a Cronin--like
enhancement in $d+Au$ collisions at midrapidity, going to a strong
suppression at $\eta \gtrsim 2$. In central $Au+Au$ collisions we find
a suppression both at mid-- and forward rapidities that vanishes for
peripheral collisions. We interpret this as signs of several
different medium related effects modifying bulk particle production in
$Au+Au$ and $d+Au$ collisions at RHIC energies.

\end{abstract}

\maketitle


The processes that govern bulk particle production in
ultrarelativistic nucleus--nucleus collisions~\cite{Wang:1991ht} are
not well understood. Exploring this experimentally is one of the goals
of the RHIC accelerator at Brookhaven National Lab., and of the BRAHMS
experiment in particular. RHIC has now collided gold nuclei on gold
($Au+Au$), protons on protons ($p+p$) and deuterons on gold ($d+Au$)
at the same center--of--mass energy of
$\sqrt{s_{NN}}=\unit[200]{GeV}$, allowing us to compare the complex
dynamics of a heavy--ion collision to the comparatively simple case of
a nucleon--nucleon interaction, and the particle production mechanisms
in hot nuclear matter (central $Au+Au$) to those in a cold nuclear
medium ($d+Au$).

The hope is that clear, systematic effects will emerge from these
comparisons that will tell us about the gross properties of the system
created in these collisions. Below we present BRAHMS data on charged
hadron production from all three colliding systems, and construct the
nuclear modification factors $R_{AuAu}$ and $R_{dAu}$ that are
sensitive to such medium--related effects. We will show that bulk
properties do indeed emerge when comparing both extended systems to
$p+p$ collisions, and indicate how this can be interpreted in terms of
several distinct mechanisms for modification of particle production in
different parts of $\eta-p_{T}$ space. Here $p_{T}$ is the particle
transverse momentum, $\eta = -\ln(\tan \frac{\theta}{2})$ its
pseudorapidity and $\theta$ the angle of the particles momentum
relative to the beamline.


\begin{figure}[htp]
  \resizebox{0.99\linewidth}{!}
  {\includegraphics{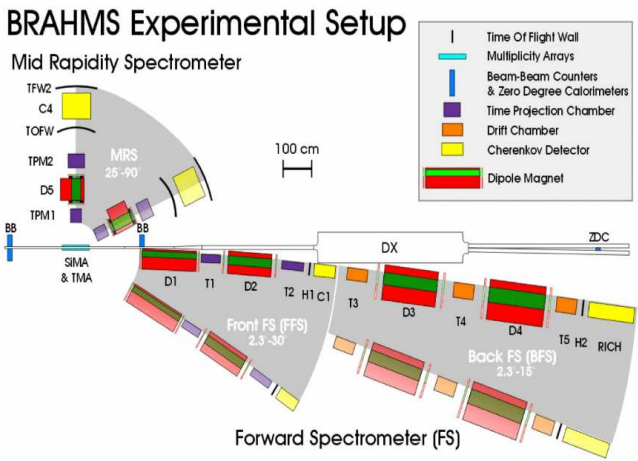}}
  \caption{The BRAHMS experiment at RHIC. For details, refer
  to~\protect{\cite{BRAHMSNIM}}.}
 \label{fig1}
 \vspace{-2mm}
\end{figure}

The BRAHMS experimental setup~\cite{BRAHMSNIM, clausbormio}, one of four detectors
at RHIC, is a two--arm hadronic magnetic spectrometer with three main
parts -- see Figure~\ref{fig1}. The first is a set of event
characterization detectors including Cherenkov scintillator arrays for
multiplicity and collision point determination (BB), silicon strips
and plastic scintillator tiles for particle multiplicity and
centrality determination (SMA \& TMA), and two zero--degree
calorimeters (ZDC) used for collision point, centrality and cross
section determination.

Secondly there is a midrapidity spectrometer arm (MRS) that can be
positioned at angles from 90 to 30 degrees with respect to the beam
axis. The MRS consists of two TPCs (TPM1, TPM2) and a dipole magnet
for tracking and momentum determination, and a time--of--flight wall
(TOFW) for particle identification. The MRS can identify $\pi^{\pm}$
and K$^{\pm}$ up to $p=\unit[2.0]{GeV/c}$ and p/$\bar{\rm p}$ up to
$p=\unit[3.0]{GeV/c}$. At a single angle and magnetic field setting
the solid angle coverage of the MRS is quite limited, but by rotating
the spectrometer and changing the field of the bending magnet a large
coverage in rapidity and transverse momentum is achieved.

Finally, the part that makes BRAHMS unique at RHIC is the forward
rapidity spectrometer arm (FS), which rotates from 20 degrees (front
part only) to 4 degrees relative to the beam axis. It consists of two
TPCs and three Drift Chambers (T1-T5), four dipole magnets (D1-D4),
one threshold and one ring--imaging Cherenkov counter (C1 and
RICH). The full FS allows identification of pions, kaons and protons
up to $\unit[20]{GeV/c}$. As for the MRS, the limited geometrical
acceptance is extended by rotating the FS and changing the fields. All
in all BRAHMS can identify pions from midrapidity and up to
$y\approx3.5$, with the RHIC beam rapidity at $y_{b}=5.3$.

For the $p+p$ and $d+Au$ data sets taken in 2003, the above setup was
augmented by two sets of plastic scintillator rings around the
beampipe (not shown in Fig.~\ref{fig1}, see Ref. \cite{Arsene:2004cn}
for a description), four rings on each side of the nominal
interaction point. These counters provided event triggering,
interaction point and cross section measurements in the relatively low
multiplicity events from these collision systems. For further details
of the BRAHMS experimental setup, see Ref.~\cite{BRAHMSNIM}. Also note
that for the $d+Au$ data presented here, the spectrometers saw only
the deuteron fragmentation side of the collision and we consequently
define this to have positive rapidity.


\begin{figure}[htp]
  \resizebox{0.99\linewidth}{!}
  {\includegraphics{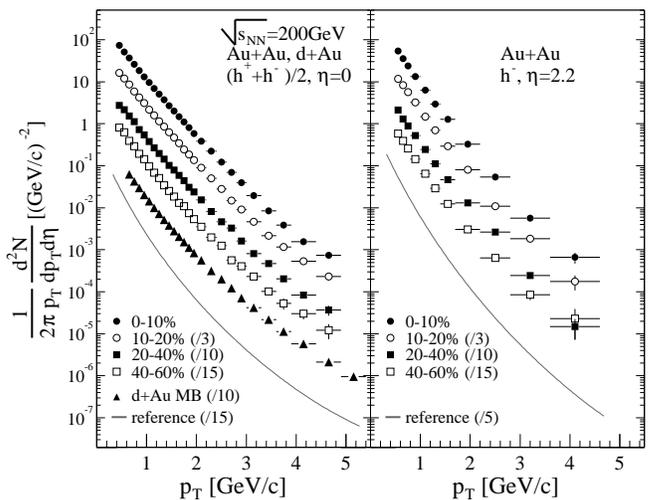}}
  \caption{Invariant spectra of charged hadrons from $Au+Au$ collisions
    at $\sqrt {s_{NN}} = \unit[200]{GeV}$ for pseudorapidities $\eta=0$ (left
    panel) and $\eta=2.2$ (right panel). Various centrality cuts are
    shown for Au+Au. The ${\rm p}+\bar{{\rm p}}$ reference spectra
    (appropriately acceptance scaled) are shown for comparison.  The
    left panel also shows the $d+Au$ spectrum.  For clarity, some
    spectra have been divided by the indicated factors. Published in~\protect{\cite{Arsene:2003yk}}.}
 \label{fig2}
\end{figure}

\begin{figure*}[!ht]
  \resizebox{0.99\textwidth}{!}
  {\includegraphics{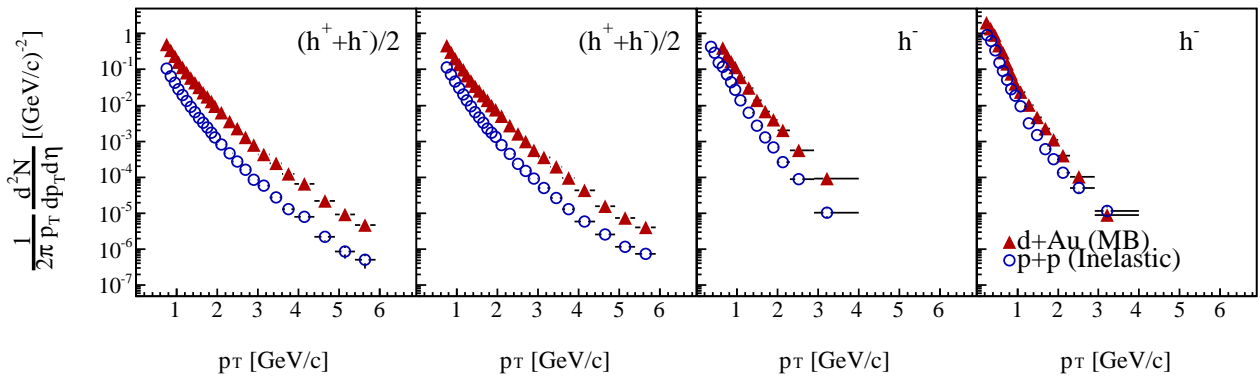}}
  \caption{\label{fig:spectra} Invariant yield
    distributions for charged hadrons produced in $d+Au$ and $p+p$
    collisions at $\sqrt{s} = \unit[200]{GeV}$ at pseudorapidities
    $\eta=0,1.0,2.2,3.2$. Published in~\protect{\cite{Arsene:2004ux}}.}
  \label{fig3}
\end{figure*}

\begin{figure}[htp]
\vspace{2mm}
  \resizebox{0.99\linewidth}{!}
  {\includegraphics{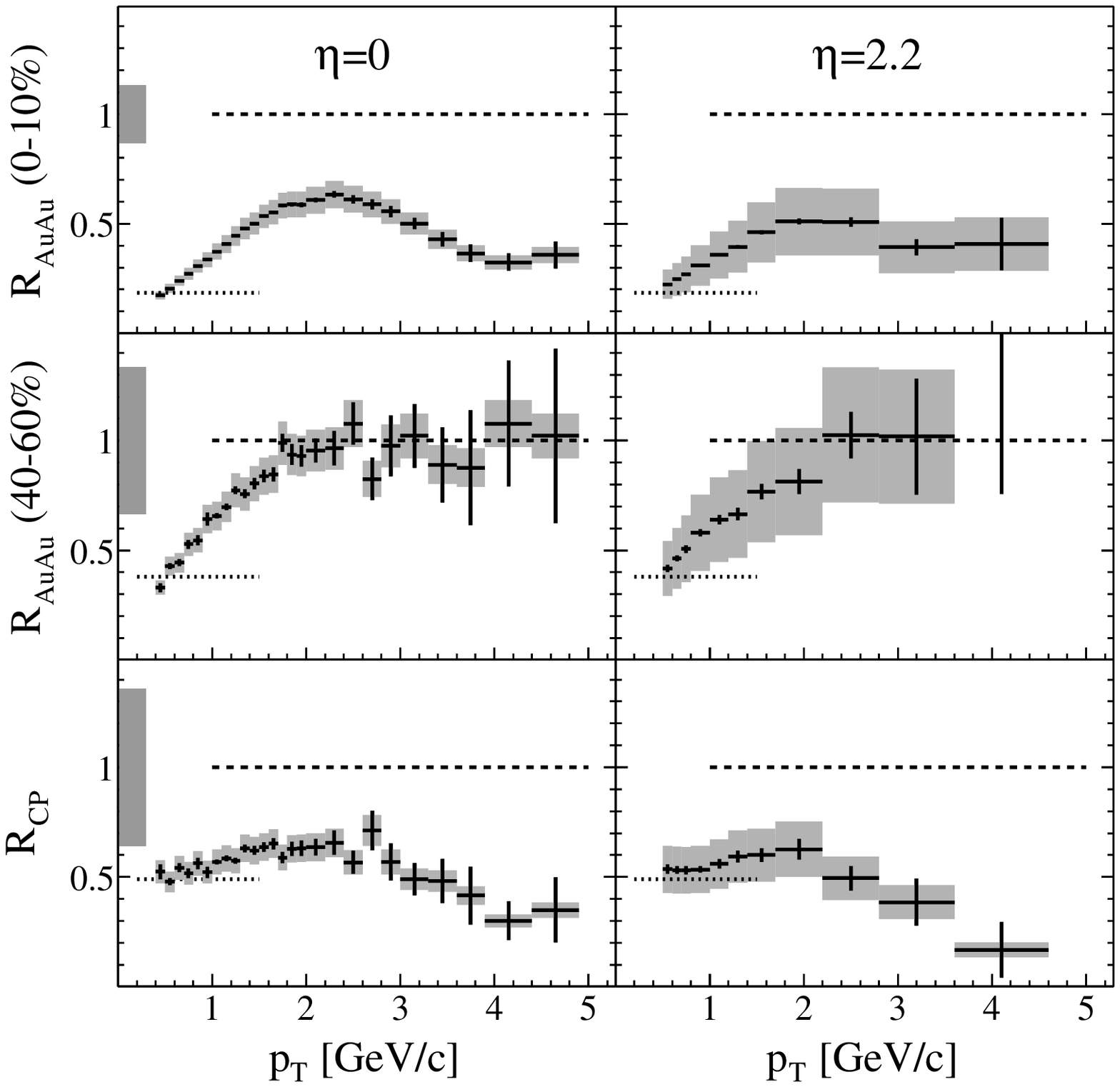}}
  \caption{ Top row: Nuclear modification factors $R_{AuAu}$ as a
    function of transverse momentum for $Au+Au$ collisions at $\eta=0$
    and $\eta=2.2$ for the $0-10\%$ most central collisions. Middle
    row: as top row, but for centralities $40-60\%$. Bottom row: ratio
    of the $R_{AuAu}$ factors for the most central and most peripheral
    collisions at the two rapidities. The dotted and dashed lines show
    the expected value of $R_{AuAu}$ using a scaling by the number of
    participants and by the number of binary collisions,
    respectively. Error bars are statistical.  The gray bands indicate
    the estimated systematic errors.  The gray band at $p_T=0$ is the
    uncertainty on the scale. Published in~\protect{\cite{Arsene:2003yk}}.}
  \label{fig4}
\end{figure}

Figures~\ref{fig2} and~\ref{fig3}~\cite{Arsene:2003yk,Arsene:2004ux}
show BRAHMS charged hadron spectra from $Au+Au$, $d+Au$ and $p+p$
collisions at $\sqrt{s_{NN}}=\unit[200]{GeV}$ for a number of
pseudorapidities and centrality classes. Measuring charged particle
production with BRAHMS is a multi--step process due to the limited
acceptance --- total $p_{T}$ spectra of charged hadrons at one
pseudorapidity \cite{Arsene:2003yk, Bearden:2003hx, Ouerdane:2002gm}
as shown are constructed by combining several such settings. Our
spectra have been corrected for finite geometrical acceptance as well
as trigger and detector efficiency using a GEANT--based Monte Carlo
simulation of the entire detector setup.  Details of the analysis can
be found in
Refs.~\cite{Arsene:2003yk,Arsene:2004ux}~\footnote{Figure~\ref{fig2}
shows a reference $p+p$ spectrum from UA(1)~\cite{UA1} scaled to our
acceptance, since this was used in determining the $R_{AuAu}$ factors
shown in Fig.~\ref{fig4}. The BRAHMS $p+p$ spectra shown in
Fig.~\ref{fig3}, determined later, are consistent with the scaled
UA(1) result.}.


\begin{figure*}[!ht] 
\resizebox{0.99\textwidth}{!}
          {\includegraphics{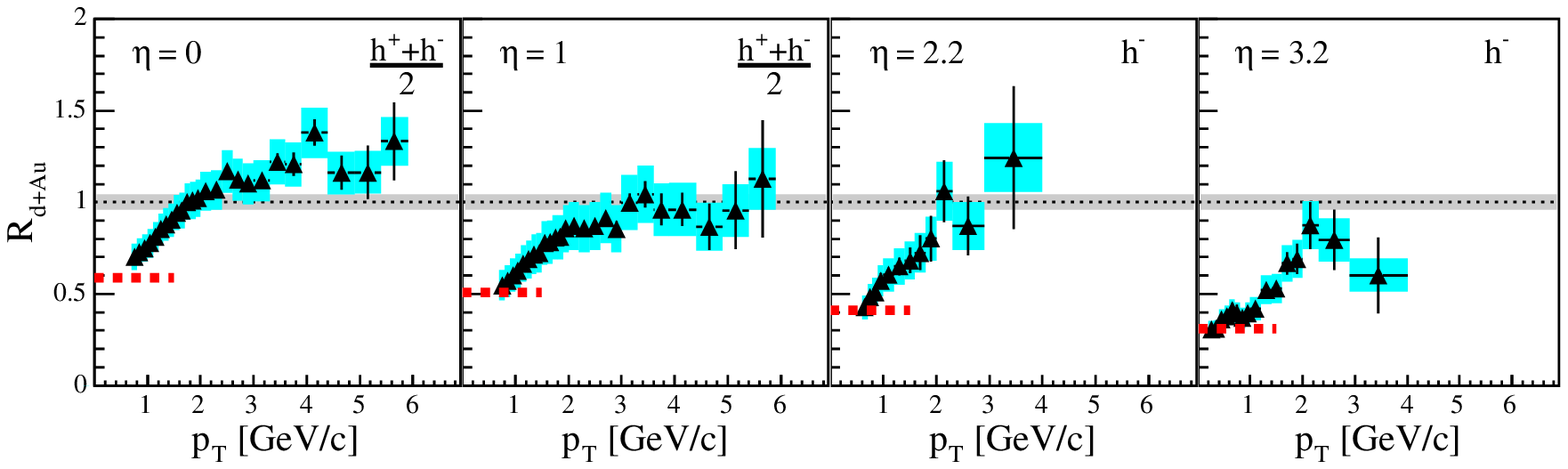}}
\caption{\label{fig:ratio}Nuclear modification factor for charged
  hadrons at pseudorapidities $\eta= 0, 1.0, 2.2, 3.2$. One standard
  deviation statistical errors are shown with error bars. Systematic
  errors are shown with shaded boxes. The vertical shaded band around
  unity indicates the error on the normalization to $\langle N_{coll}
  \rangle$. Published in~\protect{\cite{Arsene:2004ux}}.}
  \label{fig5}
\end{figure*}

Our aim is to extract any effects of a nuclear medium on the total
particle production, by comparing to elementary nucleon--nucleon
collisions where we do not expect such a medium to be produced. We
construct the Nuclear Modification Factor $R_{AB}$:
\begin{equation}
R_{AB} \equiv \frac{1}{\langle N^{AB}_{coll} \rangle}
\frac{N_{AB}(p_{T},\eta)}{N_{pp}(p_{T},\eta)}. \label{equation1}
\end{equation}
where A and B are the colliding systems (i.e. $Au$ or $d$) and
$\langle N^{AB}_{coll} \rangle$ is the average number of binary
nucleon--nucleon collisions in such an event as estimated by the
Glauber model. $N_{AB}$ and $N_{pp}$ are the charged hadron
productions in the A+B and $p+p$ collisions respectively, taken at the
same transverse momentum $p_{T}$ and pseudorapidity $\eta$.

If there were no medium effects, i.e. if a nucleus--nucleus collision
could be viewed simply as individual nucleon--nucleon collisions, then
we would expect an $R_{AB}$ factor of unity above a certain
$p_{T}$ threshold and decreasing smoothly below. The latter is because
at low $p_{T}$ we expect a scaling of particle production with the
number of participating nucleons $\langle N_{part} \rangle$ rather
than $\langle N_{coll} \rangle$. Both for $Au+Au$ and $d+Au$
collisions, we see strong deviations from this simple behavior.


In central $Au+Au$ collisions, where we expect to produce ``hot''
nuclear matter (see e.g. \cite{McLerran:2003yx}), we see a strong
suppression of particle production relative to $p+p$ scaling --- see
Fig.~\ref{fig4}~\cite{Arsene:2003yk}, upper left panel. The expected
rise at low $p_{T}$ is seen up to $p_{T} \approx \unit[2]{GeV/c}$, but then
$R_{AuAu}$ seems to level off at a value well below unity and
subsequently decrease again. This behavior is in sharp contrast to
what was seen at SPS energies, i.e. $Pb+Pb$ collisions at
center--of--mass energies of up to $\sqrt{s_{NN}}=\unit[17]{GeV}$,
where one observed a strong enhancement of $R_{PbPb}$, dubbed the
Cronin enhancement~\cite{Cronin}, interpreted as an initial state
broadening of the distribution of quark momenta ($k_{T}$ broadening). We
also see (middle left panel of Fig.~\ref{fig4}) that for more
peripheral $Au+Au$ collisions the suppression vanishes and we reach
the behavior expected from $p+p$ scaling. This indicates that we may
have an onset of a new mechanism for suppressing high--$p_{T}$
particle production between top SPS and RHIC energies, and between
peripheral and central collisions at RHIC --- i.e. in the collisions
where we reach the highest energy densities~\footnote{Using Bjorkens
estimate for the energy density $\varepsilon_{Bj} = \frac{1}{\pi
R_{A}^2 \tau} \frac{dE_{t}}{dy}$~\cite{bjorken} and RHIC measurements
of total charged particle density~\cite{Bearden:2001qq}, we find that we reach
densities of up to $\varepsilon_{Bj} \sim \unit[10-20]{GeV/fm^3}$ in
central $Au+Au$ collisions at $\sqrt{s_{NN}}=\unit[200]{GeV}$. Here,
$\tau$ is the time it takes to get to the densest state, usually set
to \unit[1-2]{fm/c}.}.  The
behavior reported here has also been seen by the other RHIC
experiments~\cite{RHIC_auau_supp}.

To extend this picture, BRAHMS has also measured $R_{AuAu}$ from
central collisions at a more forward rapidity $\eta \approx 2.2$,
shown in the upper right panel of Fig.~\ref{fig4}. The suppression
seen at midrapidity is still present here, showing that if it is due
to some final--state mechanism prevalent in the fireball, the medium
that has this suppressing property must extend out at least two units
of pseudorapidity away from $\eta=0$. Again, the suppression vanishes
for more peripheral collisions.

Using the above definition of the nuclear modification factor we need
to compare separate measurements of nucleus--nucleus and $p+p$
collisions, introducing additional systematic errors. To avoid this
problem, we can approximate $p+p$ collisions by peripheral
nucleus--nucleus collisions scaled to $1/\langle N^{Periph}_{coll}
\rangle$ and rather construct the ratio of central to peripheral
collisions $R_{CP}$:
\begin{equation}
R_{CP} \equiv \frac{1/\langle N^{Central}_{coll} \rangle}{1/\langle N^{Periph}_{coll} \rangle}
\frac{N^{Central}_{AB}(p_{T},\eta)}{N^{Periph}_{AB}(p_{T},\eta)}. \label{equation2}
\end{equation}

The bottom panels of Fig.~\ref{fig4} show the ratio $R_{CP}$ for
$Au+Au$ collisions, using the centrality classes 0-10\% and
40-60\%. Again we see a clear saturation of the modification factor at
a value below unity and a subsequent further drop as $p_{T}$
increases, showing that there is an attenuating mechanism present in
the central collisions that is either not there or greatly reduced in
peripheral collisions.


\begin{figure*} [!ht]
  \resizebox{0.99\textwidth}{!}  
  {\includegraphics{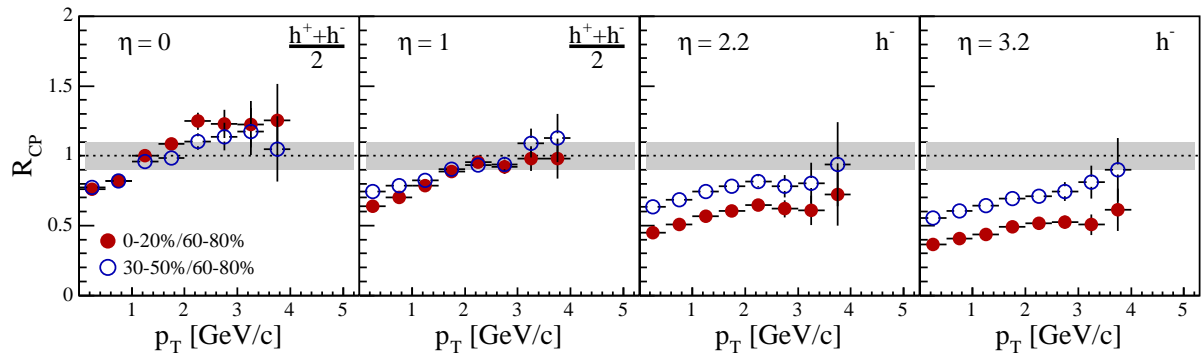}}
  \caption{\label{fig:centrality} Central (full points) and
    semi-central (open points) $R_{CP}$ ratios from $d+Au$ collisions (see text for details)
    at pseudorapidities $\eta=0, 1.0, 2.2, 3.2$. Published in~\protect{\cite{Arsene:2004ux}}.}
  \label{fig6}
\end{figure*}

Figure~\ref{fig5} shows the corresponding ratios $R_{dAu}$ for
minimum--bias $d+Au$ collisions at $\sqrt{s_{NN}}=\unit[200]{GeV}$,
for pseudorapidities $\eta = $ 0, 1.0 , 2.2 and
3.2~\cite{Arsene:2004ux}. From these collisions at midrapidity, where
the size of the fireball is smaller and we don't expect to produce an
extended hot and dense medium, we see an enhancement of $R_{dAu}$ very
similar to the Cronin effect, rather than a suppression. However as we
move to more forward rapidities, this effect vanishes and at $\eta
\gtrsim 2.2$ we once again see indications of a suppression.

In Figure~\ref{fig6}~\cite{Arsene:2004ux} we have defined three
centrality classes (0--20\%, 30--50\% and 60--80\%) and again
constructed the ratio $R_{CP}$ of central (or semi--central) to
peripheral collisions. While the factors are approximately equal at
midrapidity, there is a clear change of behavior toward higher
rapidities as the $R_{CP}$ for central collisions drops significantly
below that for semi--peripheral collisions.  This indicates that we
have an attenuating effect on forward particle production when the
deuteron hits the Lorentz contracted gold nucleus well away from the
edges, i.e. where we have a constant, dense nuclear initial state.


To interpret these observations, several distinct mechanisms may be
needed. Firstly, the midrapidity enhancement of $R_{dAu}$ can be
explained through a broadening of the initial momentum distributions
due to multiple scattering (Cronin enhancement) as for the SPS data,
indicating similar initial state medium effects at midrapidity in
$d+Au$ collisions at $\sqrt{s_{NN}}=\unit[200]{GeV}$ and in in $Pb+Pb$
collisions at $\sqrt{s_{NN}}=\unit[17]{GeV}$.

Secondly and more interestingly, the suppression of high--$p_{T}$
particle production in central $Au+Au$ collisions is indicative of
partonic energy loss due to gluonic bremsstrahlung in a colored
medium. In other words, the partons lose energy through gluonic
interactions because they traverse a medium with a high density of
color charges. The fact that the suppression disappears in more
peripheral collisions and turns to a Cronin--like enhancement for
central or min--bias $d+Au$ collisions where we do not expect to
produce a deconfined medium, indicates that this is indeed a
final--state effect and not just a consequence of the high--energy
nuclear initial state. Results from the STAR experiment on strong
suppression of far--side jets in central $Au+Au$ collisions extend and
support this conclusion~\cite{Adams:2003im}.

Thirdly, the transition from a Cronin--like enhancement at midrapidity
to a clear suppression with increasing rapidity in $d+Au$ collisions
indicates yet another mechanism, related to the nuclear initial
state. Recent theoretical work has predicted and effect similar to
this on the basis of gluon--saturation phenomena, a scenario known as
the Color Glass Condensate~\cite{McLerranVenu,KKT,Iancu:2003xm}. It is based on the
observation from deep--inelastic scattering at HERA~\cite{HERA} that
as the momentum transfer between the electron and the struck parton
increases, the observed gluon density function seems to diverge at
small values of $x$, i.e. there is a high density of so--called ``wee
gluons'' carrying a very small fraction $x$ of the total nucleon
momentum.  Due to the finite size of a nucleon, one can assume that at
some density these small--$x$ gluons will start to fuse due to gluonic
self--interaction. For a nuclear system at high energies, one can then
predict a transition to a condensed state of ``colored glass'' --- a
state with a high density of color charges that evolves slowly
compared to the timescale of an ultrarelativistic heavy--ion collision
($\approx 1fm/c$), and with the gluonic wavefunctions extended in the
transverse directions. The transverse momentum scale $Q^2_s$ for the
onset of this saturation in a nuclear collision can be shown to depend
on the gluon density and thus on the number of participating nucleons,
and is connected with the rapidity of measured particles by $Q^2_s
\sim A^{\frac{1}{3}} e^{\lambda y}$ (see e.g.~\cite{Iancu:2003xm}),
where $\lambda \sim 0.2-0.3$ is obtained from fits to HERA data. At
RHIC and for the $p_{T}$ range covered by BRAHMS, the onset of
saturation is predicted~\cite{KKT} to occur between rapidities 0 and
3, in agreement with our experimental observations.

However, in this part of phase space nuclear
shadowing~\cite{Wang:1991ht} is also important. The shadowing effect,
observed e.g. in muon--nucleus interactions~\cite{Ashman:1988bf}, is a
depletion of low--$x$ partons in nuclear systems. It has been
interpreted both as destructive interference reducing the probability
of interactions with partons at the back of an incoming nucleus, and
as gluon recombination at high densities but without the condensation
mechanism discussed above. This effect is certainly present in our
data, as shown by the observation that simulations using the HIJING
event generator which includes shadowing well reproduces BRAHMS data
on total charged particle production at all accessible
pseudorapidities~\cite{Arsene:2004cn}. A recent discussion in
Ref.~\cite{Accardi:2004ut} indicates that such effects may not be
enough to describe the results presented here, but this discussion
relies on extrapolated fits to RHIC midrapidity data. More data at
forward rapidities at higher $p_{T}$ and subsequent theoretical work
is needed to determine whether the observed suppression is indeed
partly due to the formation of a Color Glass Condensate in the initial
state at RHIC. Also, a better treatment of soft beam fragmentation
effects may be needed to properly construct the $R_{dAu}$ factor.

An effect that may affect the data and that has not been discussed in
this contribution is a possible isospin dependent contribution to the
nuclear modification factors. A nucleus has both protons and neutrons,
and so the denominator in the $R_{AB}$ factor should be a properly
weighted average of $p+p$, $p+n$ and $n+n$ collisions. Data on this at
SPS energies have been taken by the NA49 collaboration, but are
unfortunately not yet published. However, while this may have an
effect on the exact shapes of the $R_{AB}$ factors, the consequences
should not be large enough to affect the overall trends presented
above.

In conclusion, BRAHMS has measured the charged hadron production from
$Au+Au$, $d+Au$ and $p+p$ collisions at the same center--of--mass
energy of $\sqrt{s_{NN}}=\unit[200]{GeV}$, both as a function of
pseudorapidity and centrality. Our measurements cover an extensive
part of phase space, uniquely including the forward rapidity region which
probes the small--$x$ part of the nuclear structure functions. 
Through the nuclear modification factors $R_{AuAu}$ and $R_{dAu}$ we
have uncovered a suppression of high--$p_{T}$ particle production in
$d+Au$ collisions at forward rapidities, in contrast to the
enhancement previously seen at midrapidity. We also find a suppression
of $R_{AuAu}$ at high--$p_{T}$ in central collisions, extending at
least two units of pseudorapidity away from $\eta=0$. The results
presented here can be interpreted in terms of a number of
medium--related effects modifying bulk particle production:
\begin{itemize}
\item A Cronin--like enhancement in $d+Au$ collisions at midrapidity,
  due to initial state broadening of intrinsic quark transverse momenta.
\item A final--state suppression of high--$p_{T}$ particle production
  in central $Au+Au$ collisions, interpreted as a partonic energy loss
  due to gluon bremsstrahlung in a deconfined medium with a high
  density of color charges.
\item An initial state suppression of $R_{dAu}$ at forward rapidities,
  at least partly due to nuclear shadowing effects depleting the
  number of small--$x$ partons available for interaction. There may
  also be an additional suppressing effect from the saturation and
  condensation of small--$x$ gluons, as predicted in the Color Glass
  Condensate scenario.
\end{itemize}

\begin{acknowledgments}
This work was supported by the division of Nuclear Physics of the
Office of Science of the U.S. DOE, the Danish Natural Science Research
Council, the Research Council of Norway, the Polish State Com. for
Scientific Research and the Romanian Ministry of Research.

\end{acknowledgments}

\bibliography{apssamp}

\begin{thebibliography}{99}

\bibitem{Wang:1991ht}
X.~N.~Wang and M.~Gyulassy,
Phys.\ Rev.\ D {\bf 44}, 3501 (1991).

\bibitem{BRAHMSNIM} M.\ Adamczyk {\it et al.}, BRAHMS Collaboration, Nuclear
  Instruments and Methods, {\bf A499} 437 (2003).

\bibitem{clausbormio} C.~E.~J\o rgensen {\it et al.}, BRAHMS
  Collaboration, elsewhere in this proceeding.

\bibitem{Arsene:2004cn}
I.~Arsene  {\it et al.}, BRAHMS Collaboration,
arXiv:nucl-ex/0401025. Subm. to Phys. Rev. Lett.

\bibitem{Arsene:2003yk}
I.~Arsene {\it et al.}, BRAHMS Collaboration,
Phys.\ Rev.\ Lett.\  {\bf 91}, 072305 (2003), nucl-ex/0307003.

\bibitem{Arsene:2004ux}
I.~Arsene  {\it et al.}, BRAHMS Collaboration,
arXiv:nucl-ex/0403005. Subm. to Phys. Rev. Lett.

\bibitem{Bearden:2003hx}
I.~G.~Bearden {\it et al.}, BRAHMS Collaboration,
arXiv:nucl-ex/0312023. Subm. to Phys. Rev. Lett.

\bibitem{Ouerdane:2002gm}
D.~Ouerdane {\it et al.}, BRAHMS Collaboration,
Nucl.\ Phys.\ A {\bf 715}, 478 (2003), nucl-ex/0212001.

\bibitem{McLerran:2003yx}
L.~McLerran,
arXiv:hep-ph/0311028.

\bibitem{Cronin} D. Antreasyan {\it et al.}, Phys. Rev. D {\bf 19}, 764 (1979). 

\bibitem{RHIC_auau_supp}
  K. Adcox {\it et. al.}, PHENIX Collaboration, Phys. Rev. Lett. {\bf 88} 022301 (2002);
  S. S. Adler {\it et. al.}, STAR Collaboration, Phys. Rev. Lett. {\bf
  89} 202301 (2002);
  B.B. Back, {\it et. al.}, PHOBOS Collaboration, Phys.Lett. B578,
  {\bf 297} (2004) 

\bibitem{Adams:2003im}
J.~Adams {\it et al.}, STAR Collaboration,
Phys.\ Rev.\ Lett.\  {\bf 91}, 072304 (2003), nucl-ex/0306024.

\bibitem{McLerranVenu}
  L. McLerran and R. Venugopalan,
  Phys. Rev. D {\bf 49}, 2233(1994),
  Phys. Rev. D {\bf 49}, 3352 (1994),
  Phys. Rev D {\bf 50}, 2225 (1994),
  Phys. Rev. D {\bf 59}, 094002 (1999);
  Y. V. Kovchegov,
  Phys. Rev. D {\bf 54}, 5463 (1996),
  Phys. Rev. D {\bf 55}, 5445 (1997).

\bibitem{Iancu:2003xm}
E.~Iancu and R.~Venugopalan,
arXiv:hep-ph/0303204.

\bibitem{KKT} D. Kharzeev, Y. V. Kovchegov and K. Tuchin
  Phys. Rev. D {\bf 68}, 094013, (2003), hep-ph/0307037

\bibitem{HERA}
  J. Breitweg {\it et al.} Eur. Phys. J. {\bf C7} 609-630, (1999);
  ZEUS Collaboration, J. Breitweg {\it et al.}, Phys. Lett. {\bf B487} (2000) 53;
  ZEUS Collaboration, S. Chekanov {\it et al.}, Eur. Phys.J. {\bf C21} (2001) 443;
  H1 Collaboration, C. Adloff {\it et al.}, Eur. Phys. J. {\bf C21} (2001) 33.

\bibitem{Ashman:1988bf}
J.~Ashman {\it et al.}, European Muon Collaboration (EMC),
Phys.\ Lett.\ B {\bf 202}, 603 (1988).

\bibitem{Accardi:2004ut}
A.~Accardi and M.~Gyulassy,
nucl-th/0402101.

\bibitem{UA1} C. Albajar {\it et al.} Nucl. Phys. B{\bf 355} 261 (1990).

\bibitem{bjorken}
J.~D.~Bjorken,
Phys.\ Rev.\ D {\bf 27}, 140 (1983).

\bibitem{Bearden:2001qq}
I.~G.~Bearden {\it et al.}, BRAHMS Collaboration,
Phys.\ Rev.\ Lett.\  {\bf 88}, 202301 (2002), nucl-ex/0112001.


\end{thebibliography}

\end{document}